\documentstyle[prl,aps]{revtex}
\def\ie{{\it i.e.},\ }

\font\nrm=cmr9

\newcommand{\ket}[1]{\left|#1\right\rangle}

\begin{document}
\addtolength{\textwidth}{6pt}
\draft
\twocolumn[\hsize\textwidth\columnwidth\hsize\csname@twocolumnfalse%
\endcsname
\phantom{Reply to Bernevig, Giuliano, and Laughlin}
]

\noindent {\bf Comment on ``Spinon 
Attraction in Spin-1/2 Antiferromagnetic Chains''}\\[-5pt]

In a recent Letter\cite{BernevigGiulianoLaughlin01prl}, Bernevig,
Giuliano, and Laughlin (BGL) conclude that there is an attractive
interaction between spinons in the Haldane--Shastry model
(HSM)~\cite{Haldane88}.  We wish to point out here that this
conclusion is incorrect.  Spinons in the HSM actually constitute an
ideal (or free) gas of half fermions, as asserted by
Haldane~\cite{Haldane91prl2} and in some sense confirmed by
Essler~\cite{Essler95}, who calculated the spinon scattering matrix
using the asymptotic Bethe Ansatz (ABA) and found $S=\pm i$.

BGL attribute the apparent contradiction between their conclusion and
the ABA result to the fact that ``the interaction between spinons is
encoded in the definition of the pseudomomenta'' which label the Bethe
Ansatz solutions.  In other words, they assert that it is a special
feature of the ABA technique that the spinon excitations appear to be
free, while there is in fact an attractive interaction between them.
This line of thought, however, is unsustainable.  The pseudomomenta in
the ABA solutions label the exact eigenstates of the Hamiltonian.  If
the spinon scattering matrix $S$ does not depend on them, it does not
depend on the true and physical spinon momenta either, whatever they
may be.  So if the ABA is applicable to the HSM at all, which is not
garanteed {\it a priori} as the spin-spin interaction is long-ranged,
$S=\pm i$ directly and unambigously implies that the spinons are
non-interacting particles with half-fermi statistics.

Let us now critically re-examine the arguements presented by BGL.  The
first argument they give in favor of a spinon interaction is based on
a plot of $|p_{mn}(e^{i\theta})|^2$ as defined in (18) (the equation
numbers here and below refer to \cite{BernevigGiulianoLaughlin01prl})
for $m=M$, $n=0$, as a function of $\theta$.  BGL interpret
$|p_{mn}(e^{i\theta})|^2$ as probability for finding the spinons at a
distance $\theta$ along the circle from each other, and show it to be
strongly enhanced at small $\theta$.  The problem with the argument is
that, as one can easily see from (18), the
$p_{mn}(\eta_{\alpha-\beta})$'s are the coefficients in the expansion
of the overcomplete basis states $\ket{\Psi_{\alpha\beta}}$ at fixed
$\alpha$, $\beta$ in terms of $\ket{\Phi_{mn}}$.  Due to this
overcompleteness, the $p_{mn}(\eta_{\alpha-\beta})$'s as functions of
$\eta_{\alpha-\beta}$ have no direct physical
interpretation~\cite{greiterschuricht04}.

The second argument of BGL is that the last term in their expression
(10) for the energy of the two-spinon state (8) represents ``a
negative interaction contribution that becomes negligibly small in the
thermodynamic limit''.  The problem here is that BGL have identified
the momenta $q_m$ and $q_n$ of the individual spinons according to
\begin{displaymath}
  q_{m}=\frac{\pi}{2}-\frac{2\pi}{N}\!\left(m+\frac{1}{2}\right)\!,\ \ 
  q_{n}=\frac{\pi}{2}-\frac{2\pi}{N}\!\left(n+\frac{1}{2}\right)\!,  
\end{displaymath}
when interpreting the two preceding terms in (10) as the kinetic
energies of the individual spinons.  The correct identification of the
spinon momenta for $m\ge n$, however, is
\begin{displaymath}
  q_m=\frac{\pi}{2}-\frac{2\pi}{N}\!\left(m+\frac{3}{4}\right)\!,\ \
  q_n=\frac{\pi}{2}-\frac{2\pi}{N}\!\left(n+\frac{1}{4}\right)\!,
\end{displaymath}
which implies that the kinetic energy of the spinons is 
given by all three terms in the square bracket in (10).  With
$E(q)=\frac{J}{2}\bigl[\bigl(\frac{\pi}{2}\bigr)^2-q^2\bigr]$,
one finds
\begin{displaymath}
  E_{mn}=-J\frac{\pi^2}{24}\!\left(N\!+\!\frac{5}{N}\!-\!\frac{6}{N^2}\right)
  +E(q_m)+E(q_n).
\end{displaymath}
The alleged spinon interaction term has disappeared.  Physically, the
relative shift between $q_m$ and $q_n$ by one-half of a momentum
spacing $\frac{2\pi}{N}$ is a manifestation of the
half-fermi statistics of the spinons.  While the allowed values for
the total momentum $q_m+q_n$ are those for PBCs, the allowed values for
the difference in the momentum $q_m-q_n$ are those for anti-PBCs, \ie
PBCs with the ring threaded by a flux $\pi$.

Finally, BGL claim to prove through a derivation of the
Haldane--Zirnbauer formula that the spinon attraction, or more
specifically the enhancement of $|p_{mn}(e^{i\theta})|^2$ they find
when plotting it as a function of the spinon separation $\theta$, is
responsible for the square-root singularity in the dynamical spin
suceptibility (DSS).  Their reasoning is problematic in several
regards.  First, the coefficients $|p_{mn}(e^{i\theta})|^2$ cannot be
interpreted as a probability as a function of the spinon separation
$\theta$, as explained above.  Second, it is not $p_{mn}(e^{i\theta})$
as a function of $\theta$ for fixed $m$ and $n$ which enters their
derivation of the DSS, but $p_{mn}(1)$ as a function of $m$ and $n$,
as can be seen directly from (25).

In response to our considerations, BGL have countered that they merely
investigated ``the short distance effects'' of the ``statistical
interaction'' associated with the half-fermi statistics of the
spinons.  This is definitely not consistent with the claims they make
in their Letter~\cite{BernevigGiulianoLaughlin01prl}.

A more elaborate account of our understanding is given 
elsewhere~\cite{greiterschuricht04}.

\vspace{10pt}
\noindent\hbox{Martin Greiter and Dirk Schuricht\hfill}

\medskip
\vbox{
\obeylines\nrm 
Institut f\"ur Theorie der Kondensierten Materie
Universit\"at Karlsruhe
Postfach 6980
D-76128 Karlsruhe}

\vspace{5pt}
\noindent\hbox{\nrm March 30, 2005\hfill}

\noindent\hbox{\nrm PACS numbers: 75.10.Pq, 02.30.Ik, 75.10.Jm, 75.50.Ee\hfill}

\end{document}